\documentclass[8.5pt,twoside,twocolumn]{article}
\usepackage[super,sort&compress,comma]{natbib} 
\usepackage{mhchem}
\usepackage{times,mathptmx}
\usepackage{sectsty}
\usepackage{balance} 
\usepackage{color}
\usepackage{graphicx} 
\usepackage{lastpage}
\usepackage[format=plain,justification=raggedright,singlelinecheck=false,font=small,labelfont=bf,labelsep=space]{caption} 
\usepackage{fancyhdr}
\usepackage{amsmath,amssymb}
\begin{document}

\thispagestyle{plain}
\fancypagestyle{plain}{
\fancyhead[L]{\includegraphics[height=8pt]{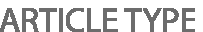}}
\fancyhead[C]{\hspace{-1cm}\includegraphics[height=20pt]{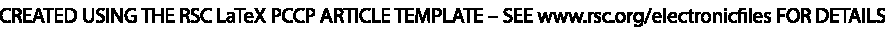}}
\fancyhead[R]{\includegraphics[height=10pt]{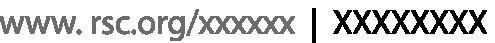}\vspace{-0.2cm}}
\renewcommand{\headrulewidth}{1pt}}
\renewcommand{\thefootnote}{\fnsymbol{footnote}}
\renewcommand\footnoterule{\vspace*{1pt}%
\hrule width 3.4in height 0.4pt \vspace*{5pt}} 
\setcounter{secnumdepth}{5}

\makeatletter 
\def\subsubsection{\@startsection{subsubsection}{3}{10pt}{-1.25ex plus -1ex minus -.1ex}{0ex plus 0ex}{\normalsize\bf}} 
\def\paragraph{\@startsection{paragraph}{4}{10pt}{-1.25ex plus -1ex minus -.1ex}{0ex plus 0ex}{\normalsize\textit}} 
\renewcommand\@biblabel[1]{#1}            
\renewcommand\@makefntext[1]%
{\noindent\makebox[0pt][r]{\@thefnmark\,}#1}
\makeatother 
\renewcommand{\figurename}{\small{Fig.}~}
\sectionfont{\large}
\subsectionfont{\normalsize} 

\fancyfoot{}
\fancyfoot[LO,RE]{\vspace{-7pt}\includegraphics[height=9pt]{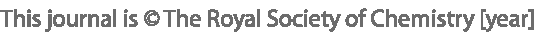}}
\fancyfoot[CO]{\vspace{-7.2pt}\hspace{12.2cm}\includegraphics{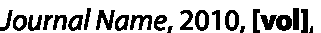}}
\fancyfoot[CE]{\vspace{-7.5pt}\hspace{-13.5cm}\includegraphics{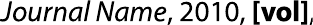}}
\fancyfoot[RO]{\footnotesize{\sffamily{1--\pageref{LastPage} ~\textbar  \hspace{2pt}\thepage}}}
\fancyfoot[LE]{\footnotesize{\sffamily{\thepage~\textbar\hspace{3.45cm} 1--\pageref{LastPage}}}}
\fancyhead{}
\renewcommand{\headrulewidth}{1pt} 
\renewcommand{\footrulewidth}{1pt}
\setlength{\arrayrulewidth}{1pt}
\setlength{\columnsep}{6.5mm}
\setlength\bibsep{1pt}

\twocolumn[
  \begin{@twocolumnfalse}
\noindent\LARGE{\textbf{Droplet-Induced Budding Transitions of Membranes}}
\vspace{0.6cm}

\noindent\large{\textbf{Halim Kusumaatmaja,$^{\ast}$\textit{$^{a}$} and Reinhard Lipowsky\textit{$^{a}$}}}\vspace{0.5cm}

\noindent\textit{\small{\textbf{Received Xth XXXXXXXXXX 20XX, Accepted Xth XXXXXXXXX 20XX\newline
First published on the web Xth XXXXXXXXXX 200X}}}

\noindent \textbf{\small{DOI: 10.1039/b000000x}}
\vspace{0.6cm}

\noindent \normalsize{Motivated by recent experiments on biomimetic membranes exposed to several aqueous phases, we theoretically study the morphology of a membrane in contact with a liquid droplet formed via aqueous phase separation. We concentrate on membranes with negligible spontaneous curvature. At small droplet volume, bending energy dominates and the droplet is only partially wrapped by the membrane. At large volume, this configuration can become unstable and undergo a discontinuous transition to a state, in which the droplet is (almost) completely wrapped by the membrane. A morphology diagram, showing the parameter region where such budding transition occurs, is constructed as a function of the membrane tension and the intrinsic contact angle of the liquid with the membrane. The effects of spontaneous curvature are discussed qualitatively.}
\vspace{0.5cm}
 \end{@twocolumnfalse}
  ]

\section{Introduction}

\footnotetext{\textit{$^{a}$~Theory \& Biosytems, Max Planck Institute of Colloids and Interfaces, 14424 Potsdam, Germany. E-mail (HK): halim@mpikg.mpg.de. E-mail (RL): lipowsky@mpikg.mpg.de.}}


Recently biomimetic membranes exposed to several aqueous phases have been introduced experimentally \cite{Li, Long, LiPNAS, Dominak}. They are found to exhibit a number of interesting and surprising phenomena, such as partial to complete wetting transitions \cite{Li}, budding \cite{Long}, and membrane tube formation \cite{LiPNAS}, which are not yet fully understood. The origin of these diverse phenomena is the competition between the bending rigidity of the membrane and the interfacial tensions of the participating phases. Essentially we are dealing with wetting phenomena on surfaces which are flexible and can attain many different morphologies. 

\begin{figure}
\centering
\includegraphics[scale=1.0,angle=0]{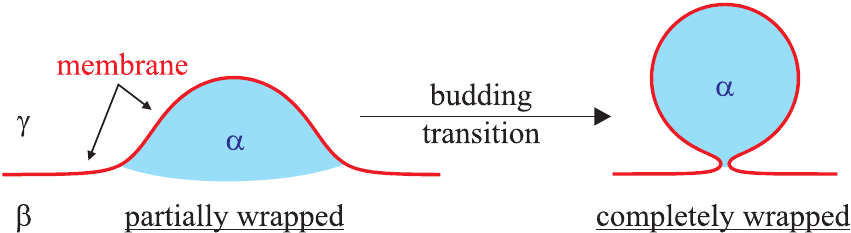}
\caption{Droplet-induced budding transition: The membrane separates the exterior aqueous phase $\gamma$ from the two coexisting interior aqueous phases $\alpha$ and $\beta$. The $\alpha$ droplet partially wets the membrane.}
\label{Fig1}
\end{figure}

A particularly interesting process we shall focus on here is droplet-induced budding which represents a morphological transition from a state where the liquid droplet is partially wrapped by the membrane to a state where the droplet is almost completely wrapped by the membrane. The resulting bud is connected to the original membrane by a small neck, as shown in Fig. \ref{Fig1}. 

Budding is an important and frequent cellular process. For example, it represents an important step during endo- and exocytosis of all membranes. Endocytosis leads to the formation of transport vesicles \cite{transport}, which allow communication and transport of biomolecules between different organelles. Likewise, budding also occurs during viral replication processes \cite{viral}. In addition, there have been proposals for using synthetic membranes for technological applications, e.g. \cite{Discher,Orwar,Libchaber}. In such a case, it is often desirable to mimic real cellular processes, including budding, as the modus operandi. This is a particularly promising approach in the context of microfluidics, the miniaturization of fluidic operations.

Given the relevance of budding for biological and biomimetic systems, it is important to understand its possible mechanisms. Budding can be induced by intramembrane domains as first predicted theoretically \cite{Lipowsky,Julicher} and confirmed experimentally by optical microscopy \cite{Baumgart,Baumg,Bacia,Dimova,Semrau}. The budding process then depends on the elastic properties of the membrane domains and on the line tension of the domain boundary. For biological membranes, the domains may contain assemblies of proteins, \cite{McMahon,Hurley} which are inserted into the membrane with a prefered orientation and, thus, induce a spontaneous curvature of the domain.

In this article, we consider a novel budding mechanism where the driving force is the interfacial tension between the aqueous phases, see Fig. \ref{Fig1}. Droplet-induced budding is somewhat similar to the encapsulation of solid or rigid particles, to which the membrane adheres \cite{Lipowsky2,Deserno,Nowak}. However, in contrast to such particles, the liquid droplets considered here change their shape during the budding process. While such a budding phenomenon has been reported experimentally \cite{Long}, there is no detailed theoretical analysis yet. It is our aim here to shed light on this new budding mechanism. In particular, we address the required recipes for budding to occur and whether the transition is continuous or discontinuous. We shall focus on the limiting, yet instructive, case, for which the bud size is small compared to the original membrane area and the membrane spontaneous curvature is negligible. Our calculations show that for a sufficiently low membrane tension, there is a critical droplet volume beyond which the partially wrapped configuration is unstable and a bud is formed. How low the membrane tension must be for a budding transition to occur depends on the wetting properties of the liquid on the membrane. This dependence can be summarized in a morphology diagram. Finally, we discuss how the membrane spontaneous curvature may modify the budding transition.

\section{Theoretical Description.}
\begin{figure}
\centering
\includegraphics[scale=1.0,angle=0]{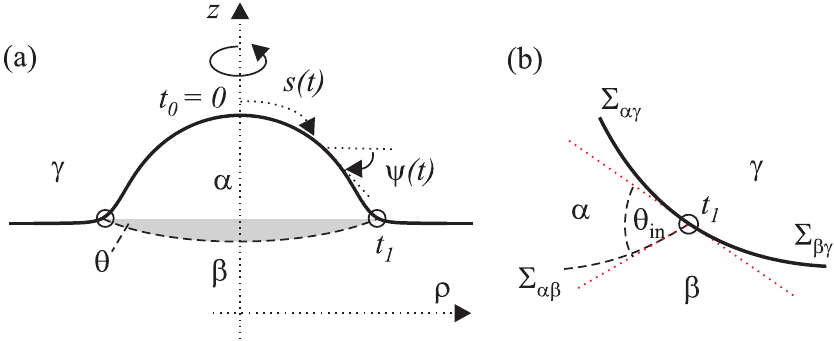}
\caption{(a) Geometry of an axisymmetric $\alpha$ droplet. The coordinate along the symmetry axis is denoted by $z$, the distance from this axis by $\rho$, the arc length by $s$, and the local tilt angle by $\psi$. All these variables depend on the contour parameter $t$. The contact line is located at $t=t_1$. The gray area corresponds to a spherical cap with tangent angle $\theta$. (b) Enlarged view close to the contact line: Intrinsic contact angle $\theta_{\mathrm{in}}$ between the two planes that are tangential to the $\alpha\beta$ interface and to the smoothly curved vesicle membrane, respectively.}
\label{Fig2}
\end{figure}
The starting point of our calculation is the theory described in \cite{Kusumaatmaja} for membranes and vesicles in contact with two aqueous phases. Here, we consider the geometry shown in Fig. \ref{Fig2}. The total energy $E$ of the latter system has several contributions. First, it contains a term that depends on the volume of the liquid droplet $V_{\alpha}$ and its pressure difference $P_{\gamma\alpha} \equiv P_{\gamma}-P_{\alpha}$ with the exterior phase $\gamma$. The aqueous phase $\beta$ is taken to be much larger than the droplet $\alpha$, which implies $P_{\gamma}-P_{\beta} = 0$, and the $\beta$ phase volume contribution can be ignored. Second, the $\alpha\beta$ interface with area $A_{\alpha\beta}$ and interfacial tension $\Sigma_{\alpha\beta}$, as well as the two membrane segments with areas $A_{\alpha\gamma}$ and $A_{\beta\gamma}$ and membrane tensions $\Sigma_{\alpha\gamma}$ and $\Sigma_{\beta\gamma}$ contribute three surface terms. The membrane tensions  $\Sigma_{\alpha\gamma}$ and $\Sigma_{\beta\gamma}$ are mechanical tensions that can be related to the anisotropic pressure tensor across the membrane. Thus, the tension $\Sigma_{\beta \gamma}$ is well-defined and remains, in general, nonzero in the limit in which the two aqueous phases $\beta$ and $\gamma$ become identical, see \cite{Goetz}. In principle, the contact line provides an energetic contribution proportional to its length $L_{\alpha\beta\gamma}$ and to its line tension $\lambda$, \cite{Kusumaatmaja} but for simplicity we have neglected this contribution. Finally, the vesicle energy contains the bending energies $E_{\mathrm{be},i}$ of the two membrane segments, with $i = \alpha,\beta$. The bending rigidity, the mean curvature, and the spontaneous curvature of the membrane are denoted by $\kappa$, $M$ and $m$ respectively. We further assume that the bending rigidities and sponteneous curvatures are the same in both membrane segments. The Gaussian curvature term is neglected since we will not consider any topological change. Thus, the total energy has the form 
\begin{eqnarray}
E = \sum_{i=\alpha,\beta} \int dA_{i\gamma} \left[2\kappa (M-m)^2 + \Sigma_{i\gamma}\right] + \Sigma_{\alpha\beta}A_{\alpha\beta} + P_{\gamma\alpha}V_{\alpha}. \label{energy}
\end{eqnarray}

As shown in Ref. \cite{Kusumaatmaja}, the force balance along the contact line is characterized by the {\it{intrinsic}} contact angle  $\theta_{\mathrm{in}}$ of the liquids at the membrane which satisfies
\begin{equation}
\frac{\Sigma_{\beta\gamma}-\Sigma_{\alpha\gamma}}{\Sigma_{\alpha\beta}} = \cos{\theta_{\mathrm{in}}},\label{intrinsic}
\end{equation}
and represents a {\it{hidden}} material property of the system. The definition of the intrinsic contact angle is shown schematically in Fig. \ref{Fig2}(b). It represents the contact angle at the nanometer scale between the $\alpha\beta$ interface and the membrane surface. Substituting Eq. (\ref{intrinsic}) into (\ref{energy}) and defining the rescaled areas $\bar{A}_{i\gamma} \equiv A_{i\gamma}(\Sigma_{\alpha\beta}/\kappa)$ and $\bar{A}_{\alpha\beta} \equiv A_{\alpha\beta}(\Sigma_{\alpha\beta}/\kappa)$, volume $\bar{V}_\alpha \equiv V_\alpha(\Sigma_{\alpha\beta}/\kappa)^{3/2}$, membrane curvature $\bar{M} \equiv M(\kappa/\Sigma_{\alpha\beta})^{1/2}$, spontaneous curvature $\bar{m} \equiv m(\kappa/\Sigma_{\alpha\beta})^{1/2}$, pressure $\bar{P}_{\gamma\alpha} \equiv P_{\gamma\alpha}(\kappa/\Sigma^3_{\alpha\beta})^{1/2}$, and membrane tension $\bar{\Sigma}_{i\gamma} \equiv \Sigma_{i\gamma}/\Sigma_{\alpha\beta}$, we obtain the rescaled energy
\begin{eqnarray}
\bar{E} \equiv E/\kappa &=& \sum_{i=\alpha,\beta} \int d\bar{A}_{i\gamma} 2(\bar{M}-\bar{m})^2 + \cos{\theta_{\mathrm{in}}}\bar{A}_{\alpha\gamma}+ \bar{\Sigma}_{\beta\gamma}\bar{A}_{\mathrm{me}} \nonumber \\
&+&\bar{A}_{\alpha\beta} + \bar{P}_{\gamma\alpha}\bar{V}_{\alpha}, \label{energy2}
\end{eqnarray}
where $\bar{A}_{\mathrm{me}} = \bar{A}_{\alpha\gamma}+\bar{A}_{\beta\gamma}$ is the rescaled total membrane area. 

We consider axisymmetric shapes which we calculate using the same procedure as in \cite{Kusumaatmaja,Julicher}. We choose the symmetry axis to be the $z$-axis. The distance from this axis will be denoted by $\rho$. The vesicle shape is then uniquely described by its one-dimensional contour $\left(z(t),\rho(t)\right)$, as shown in Fig. \ref{Fig2}(a). The parameter $t$ is the contour parameter, which varies over two fixed intervals: $t_0=0\leq t \leq t_1$ and $t_1\leq t \leq t_2$, corresponding to the two membrane segments $\bar{A}_{\alpha\gamma}$ and $\bar{A}_{\beta\gamma}$. Here $t_1$ is the position of the contact line, and we will focus on the limit of large $t_2$, and thus large $\bar{A}_{\beta\gamma}$.  Two additional quantities that play an important role in the theory are the arc length $s(t)$ and the tilt angle $\psi(t)$. Using this parameterization, the total energy $\bar{E}$ of the vesicle can be written as
\begin{equation}
\frac{\bar{E}}{2\pi}= \sum_{i=\alpha,\beta} \int dt \mathcal{L}_i+\frac{1}{3}\bar{R}^2_{\alpha\beta}(1-\cos^3{\theta})\label{energy3}
\end{equation}
where the tangent angle $\theta$ corresponds to the angle between the $\alpha\beta$ interface and the horizontal plane, see Fig. \ref{Fig2}(a), the rescaled curvature radius $\bar{R}_{\alpha\beta}$ is determined by the Laplace equation $\bar{P}_{\gamma\alpha}=-2/\bar{R}_{\alpha\beta}$, and the {\it{Lagrange functions}}
\begin{eqnarray}
\mathcal{L}_i &\equiv& \frac{1}{2}\rho s'\left(\frac{\psi'}{s'}+\frac{\sin{\psi}}{\rho} - 2\bar{m}\right)^2 + \bar{\Sigma}_{i\gamma}\rho s' \nonumber \\
&+& \frac{1}{2}\bar{P}_{\gamma i}\,\rho^2 s' \sin{\psi} + \Upsilon (\rho' - s' \cos{\psi}).
\end{eqnarray}
The Lagrange multiplier $\Upsilon$ is used to ensure the geometrical relation $\rho'=s' \cos{\psi}$. The primes correspond to derivatives with respect to contour parameter $t$.

The first variation of the energy $\bar{E}$ along the membrane surface leads to the following {\it{Euler-Lagrange}} or shape equations
\begin{eqnarray}
&\ddot{\psi} = \frac{\cos{\psi}\sin{\psi}}{\rho^2} - \frac{\dot{\psi}}{\rho}\cos{\psi} + \frac{\bar{P}_{\gamma i}}{2}\rho\cos{\psi}+\frac{\Upsilon}{\rho}\sin{\psi} ,\nonumber\\
&\dot{\Upsilon} = \frac{1}{2}(\dot{\psi}-2\bar{m})^2 - \frac{\sin^2{\psi}}{2\rho^2}+\bar{\Sigma}_{i\gamma}+\bar{P}_{\gamma i}\rho\sin{\psi} , \label{Lagrange} \\
&\dot{\rho} = \cos{\psi} \nonumber,
\end{eqnarray}
while its first variation along the contact line leads to the boundary conditions
\begin{eqnarray}
\dot{\psi}(s_1)_{\beta} = \dot{\psi}(s_1)_{\alpha}, \;\;\; \ddot{\psi}(s_1)_{\beta}-\ddot{\psi}(s_1)_{\alpha} = \sin{\theta_{\mathrm{in}}}, \label{bc}
\end{eqnarray}
with the intrinsic contact angle $\theta_{\mathrm{in}}$ as given by Eq. (\ref{intrinsic}). The overdots now denote derivatives with respect to the arc length $s$, rather than the contour parameter $t$. Furthermore, we impose the condition that the membrane is essentially flat for large $t_2$. In practice, this is done by imposing $|\psi| =|\dot{\psi}| = |\ddot{\psi}| < \epsilon$ at finite, but very large $t_2$. We typically choose $\epsilon$ to be of order $10^{-2}$ and $t_2$ to be of order one million of the discretization steps. These differential equations are then solved using the standard fourth order Runge-Kutta method \cite{RungeKutta}.

\section{The budding transition.} 

In this paper, we concentrate mainly on membranes with negligible sponatenous curvature, $\bar{m} = 0$. In this case, the morphology of the system is determined by three independent dimensionless parameters: the intrinsic contact angle $\theta_{\mathrm{in}}$, the membrane tension $\bar{\Sigma}_{\beta\gamma}$ and the droplet volume $\bar{V}_{\alpha}$. We note that generally the membrane tensions $\bar{\Sigma}_{\alpha\gamma} \neq \bar{\Sigma}_{\beta\gamma}$. They are related by the equation for the intrinsic contact angle, Eq. (\ref{intrinsic}). The effects of non-zero spontaneous curvature will be briefly discussed in section 6.

\begin{figure}[t]
\centering
\includegraphics[scale=1.0,angle=0]{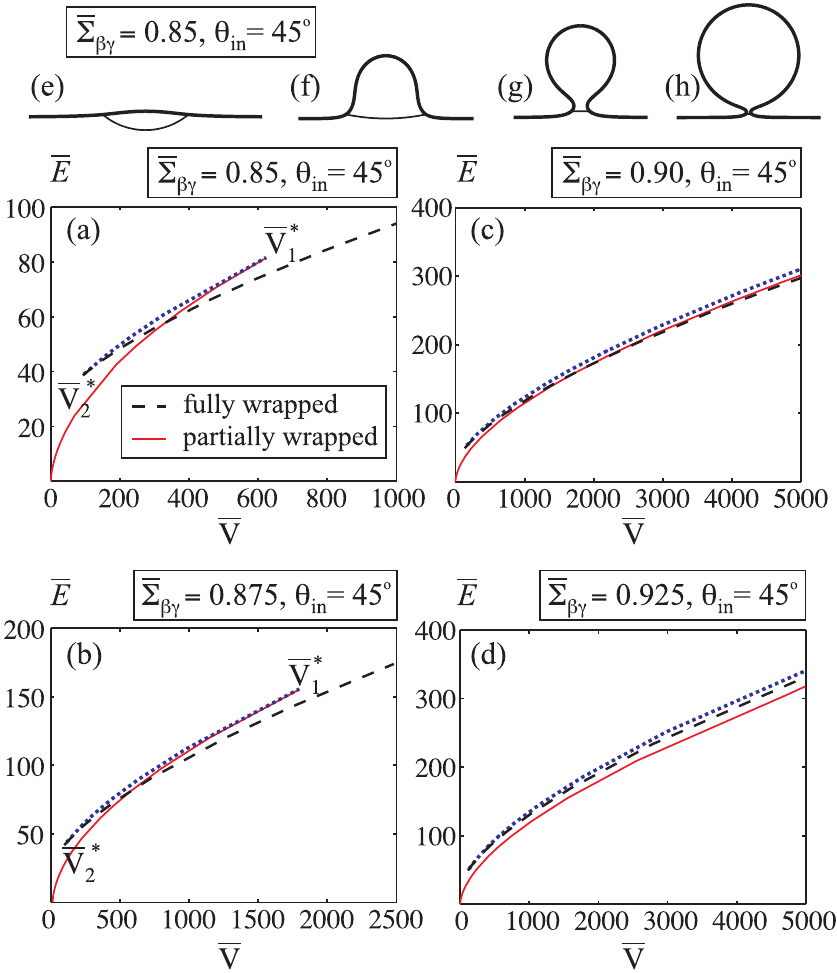}
\caption{Total energy of the system as a function of the droplet volume for various membrane tensions. At low tensions (a-b), the partially wrapped droplet (full lines) become unstable when the droplet volume $\bar{V}$ exceeds the threshold volume $\bar{V}_1^*$, whereas the completely wrapped droplet (dashed lines) becomes unstable for $\bar{V} \leq \bar{V}_2^*$. The dotted lines correspond to energy barriers. In (c), the completely wrapped configuration is only slightly more preferable to the partially wrapped. Thus, both configurations can coexist over a wide range of droplet volume. At large membrane tensions (d), the partially wrapped configuration always has a lower energy. Nonetheless, the completely wrapped configuration is still metastable. (e-g) Typical morphologies with increasing droplet volume.}
\label{Fig3}
\end{figure}

Fig. \ref{Fig3}(a) shows the typical energy curve as a function of volume for a given intrinsic contact angle and (low) membrane tension. As a representative example, we have taken $\theta_{\mathrm{in}} = 45^\circ$. In the calculations, we have also used a flat membrane with no liquid droplet as the reference surface. As mentioned, we consider the limit in which the area of this reference surface becomes large. Thus the energies shown in Fig. \ref{Fig3} are the deviations from the energy of this flat reference surface. Initially when the liquid volume is small, bending energies dominate and the membrane bends very weakly, see Fig. \ref{Fig3}(e). The system thus behaves in a similar fashion to the usual wetting geometry. With increasing liquid volume, the interfacial energies become more important and compete with the bending terms. A rather useful naive concept to have in mind is that the interfacial energy terms scale as the square of the length scale, while the bending term has no explicit length scale dependence. At higher volume, as shown in Fig. \ref{Fig3}(f), while the membrane is highly deformed, the system is still in the partially wrapped configuration. This remains the case until a certain critical volume $\bar{V}_1^*$ above which it becomes unstable and assumes the completely wrapped configuration (Fig. \ref{Fig3}(g)). From here on, increasing the liquid volume further increases the size of the spherical bud and at the same time reduces the size of the neck (Fig. \ref{Fig3}(h)).

It is important to realize that the partially and completely wrapped configurations may coexist over a range of volumes. As a result, the budding transition described above exhibits a hysteretic behaviour. The threshold volume at which the morphological instability occurs depends on whether we are increasing ($\bar{V}_1^*$) or decreasing ($\bar{V}_2^*$) the liquid volume. Between $\bar{V}_1^*$ and $\bar{V}_2^*$, there is an energy barrier for a morphological transition between partially and completely wrapped configurations. In this volume range, the shape equations (\ref{Lagrange}) and (\ref{bc}) have three solutions. The two lower energy configurations correspond to partially and completely wrapped configurations, while the third provides the energy barrier. The latter is plotted as dotted lines in Fig. \ref{Fig3}.

Increasing the membrane tension, Fig. \ref{Fig3}(b), we find that it becomes harder to bend the membrane. Thus, the budding transition occurs at a higher droplet volume. Furthermore, the volume range, for which both the partially and completely wrapped configurations coexist, increases with increasing tension. In fact, if we increase the membrane tension even further, Figs. \ref{Fig3}(c-d), we find that the two configurations effectively always coexist. In Fig. \ref{Fig3}(c), the completely wrapped configuration eventually becomes the global minimum configuration for large volumes. In contrast, the partially wrapped configuration represents the state of lowest energy for all volumes in Fig. \ref{Fig3}(d).

\begin{figure}
\centering
\includegraphics[scale=1.0,angle=0]{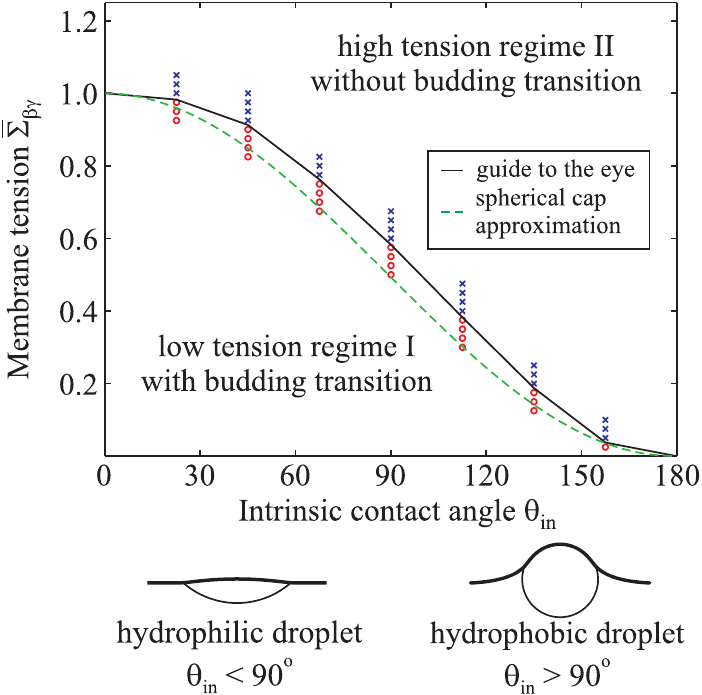}
\caption{Morphology diagram as a function of membrane tension $\bar{\Sigma}_{\beta\gamma}$ and intrinsic contact angle $\theta_{\mathrm{in}}$ with two distinct regimes: high and low tension regimes.  The data points correspond to the full numerical solutions: circles indicate the occurance of budding transitions and crosses the absence of budding transitions. We also compare the numerical results with the simple prediction (\ref{scap}), which is based on the spherical cap approximation.}
\label{Fig4}
\end{figure}

\section{Morphology diagram.}
The behaviour described in the previous section is, in fact, typical for all values of the intrinsic contact angle. Thus, we may construct a morphology diagram as a function of intrinsic contact angle and membrane tension. As shown in Fig. \ref{Fig4}, there are two regimes: (I) a low tension regime, in which the completely and partially wrapped configurations represent the state of lowest energy for large and small volumes, respectively; and (II) a high tension regime, in which the partially wrapped configuration is the state of lowest energy for all volumes. The wetting properties of the liquids on the membrane play an important role. For smaller contact angles $\theta_{\mathrm{in}}$, the liquid droplet wants to increase its contact area with the membrane, which favors the budded state. Therefore for smaller $\theta_{\mathrm{in}}$, the budding transition occurs over a larger range of membrane tension. 

We also note that, close to the boundary between regimes (I) and (II), it can be rather difficult to decide numerically if a budding transition is present. A typical example is provided by Fig. \ref{Fig3}(d). Both the partially and completely wrapped configurations are metastable over a wide range of droplet volumes, with the former having a lower energy than the latter within the computed volume range. It is, however, conceivable that if the completely wrapped configuration will become the state of lowest energy at an even larger droplet volume. In such a case, we compute the energy gradient $d\bar{E}/d\bar{V}$ for the two morphologies. If this gradient is larger for the completely wrapped configuration in the limit of large droplet volumes, then we conclude that the system will not exhibit a budding transition.

\section{Spherical cap approximation.}
\begin{figure}
\centering
\includegraphics[scale=1.0,angle=0]{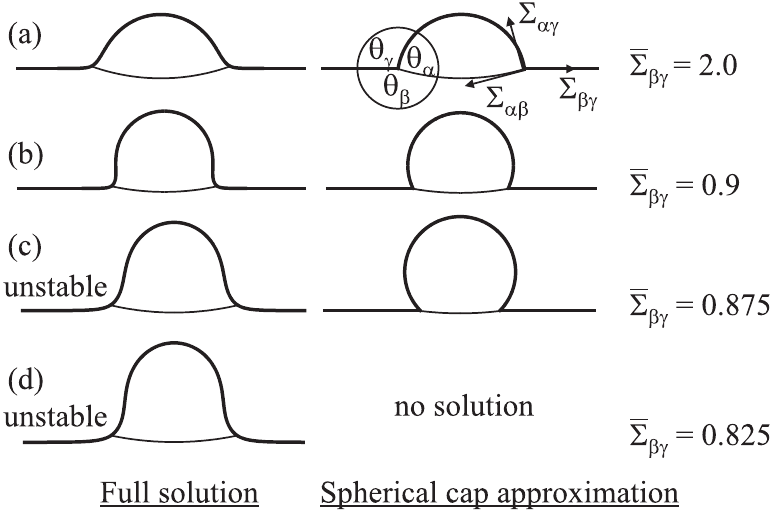}
\caption{Comparison of shapes as obtained from the full numerical solutions and from the spherical cap approximations for $\theta_{\mathrm{in}}=45^\circ$. The latter approximation breaks down at low membrane tension, but is accurate for high tension. The definitions of the tensions and effective angles are also shown.}
\label{Fig5}
\end{figure}
We shall now argue that a simple, spherical cap approximation as in Ref. \cite{Lipowsky} may be used to estimate the boundary between the two tension regimes shown in Fig. \ref{Fig4}. Such an estimate is rather useful since the full numerical solutions are time consuming to calculate. The spherical cap approximation is based on the assumption that the contributions from the interfacial terms dominate the bending terms when the volume of the droplet is large. This reflects the fact that the two energetic contributions scale differently with size. When this approximation is valid, the global shape of the system can be simply determined by considering the force balance of the interfacial tension with the membrane tensions. This is analogous to Neumann's triangle for the force balance between capillary surfaces \cite{Widom}. The resulting {\it{effective}} angles $\theta_{\alpha}$, $\theta_{\beta}$ and $\theta_{\gamma}$, defined in Fig. \ref{Fig5}, are given by
\begin{eqnarray}
\cos{\theta_i} = \frac{\Sigma^2_{jk}-\Sigma^2_{ij}-\Sigma^2_{ik}}{2\Sigma_{ij}\Sigma_{ik}}
\end{eqnarray}
with $i,j,k =  \alpha,\beta,\gamma$ and $i \neq j \neq k$, compare also to Ref. \cite{LiPNAS, Kusumaatmaja}. The sum of the effective angles is equal to $2\pi$. In contrast to capillary surfaces, for which Neumann's triangle applies, the membrane tensions, as well as the effective contact angles, do not represent material parameters; in contrast to the intrinsic contact angle as we have shown in \cite{Kusumaatmaja}. However, the three tensions are related via the relation for the intrinsic contact angle, Eq. (\ref{intrinsic}).

In Fig. \ref{Fig5}, we compare some shapes as obtained from the spherical cap approximations with the full numerical solutions. At high membrane tensions, the spherical cap model predicts the partially wrapped configuration well, see Fig. \ref{Fig5}(a). At lower membrane tensions, close to the boundary in the morphology diagram, the spherical cap approximation becomes less accurate, compare Fig. \ref{Fig5}(b), and it may give a solution corresponding to a partially wrapped configuration even when such a solution no longer exists for the original differential equations described in section 2, see Fig. \ref{Fig5}(c). Finally, at very low tensions, the spherical cap approximation does not have a solution anymore, as in Fig. \ref{Fig5}(d). Using the force balance at the contact line, the solution is lost when
\begin{equation}
\Sigma_{\alpha\gamma}+\Sigma_{\beta\gamma} < \Sigma_{\alpha\beta} 
\end{equation}
which implies
\begin{equation}
\frac{\Sigma_{\beta\gamma}}{\Sigma_{\alpha\beta}} = \bar{\Sigma}_{\beta\gamma} < \frac{1+\cos{\theta_{\mathrm{in}}}}{2}. \label{scap}
\end{equation}
This restriction does not exist in the full numerical solutions because the bending terms can compensate the force imbalance between the interfacial and membrane tensions. This importance of bending terms in the limit of infinite droplet volume implies a small contact line radius, which is consistent with a droplet in the completely wrapped configuration. Therefore, the inequality (\ref{scap}) represents a criterion for the possible existence of a completely wrapped state, and thus of a budding transition. The limiting case of relation  (\ref{scap}) as given by $\bar{\Sigma}_{\beta\gamma} = (1+\cos{\theta_{\mathrm{in}}})/2$ corresponds to the dashed line in the morphology diagram, Fig. \ref{Fig4}. Inspection of Fig. \ref{Fig4} shows that this line as obtained from the spherical cap approximation provides a rather good estimate for the boundary between the low and high tension regimes as determined by the full numerical solution. Compared to the full numerical solutions, the spherical cap approximation always predicts that the budding transition is lost at a lower membrane tension for a given intrinsic contact angle.

In this study, we have ignored thermally excited undulations, which can be justified as follows. For a membrane tension of the order of $\Sigma_{\alpha\beta}$, the longest wavelength of these undulations is about $(\kappa/\Sigma_{\alpha\beta})^{1/2}$. Using the typical experimental values\cite{Kusumaatmaja}, $\Sigma_{\alpha\beta} \sim 10^{-19}\,\mathrm{J/m^2} $ and $\kappa \sim 10^{-5}\,\mathrm{J}$, one obtains $(\kappa/\Sigma_{\alpha\beta})^{1/2}\simeq 100\,\mathrm{nm}$, which is well below optical resolution. Furthermore, the area stored in these undulations is of the order of $\frac{k_BT}{8\pi\kappa}\ln{\left[\frac{\kappa/\Sigma_{\alpha\beta}}{l^2_{\mathrm{me}}}\right]}$ times the membrane area, \cite{Lipowsky} where $l_{\mathrm{me}}$ is the small wavelength cut-off of the order of the membrane thickness. For $\kappa \simeq 10-20 \,\, k_BT$ at room temperature, the area fraction stored in thermally excited undulations is less than 1-2 percents.

\section{Spontaneous curvature.}
\begin{figure}
\centering
\includegraphics[scale=1.0,angle=0]{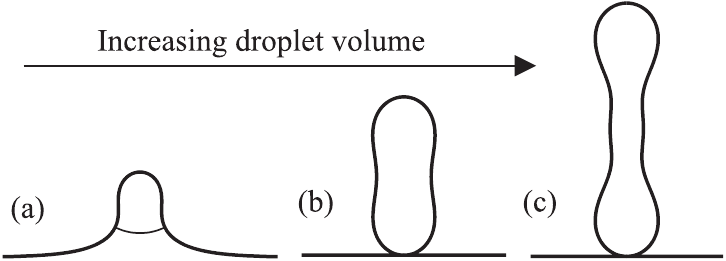}
\caption{Evolution of membrane shape with increasing droplet volume for spontaneous curvature $m = 5 \sqrt{\Sigma_{\alpha\beta}/\kappa}$, vanishing membrane tension $\bar{\Sigma}_{\beta\gamma}$, and intrinsic contact angle $\theta_{\mathrm{in}}=45^\circ$. The spontaneous curvature leads to nonspherical, tube-like membrane shapes.}
\label{Fig6}
\end{figure}
So far, we have only considered the case where the membrane spontaneous curvature has a negligible effect. In this section, we shall briefly discuss the influence of the sponteneous curvature on the budding transition shown in Fig. \ref{Fig1}. For small membrane tensions, this transition increases the bending energy of the membrane by about $8\pi\kappa$, and decreases the interfacial energy by about $\Sigma_{\alpha\beta}V_{\alpha}^{2/3}$. The budding transition occurs when these two energies are comparable, i.e. for droplet size $V_{\alpha}^{1/3} \sim (\kappa/\Sigma_{\alpha\beta})^{1/2} \simeq 100$ nm. The spontaneous curvature $m$ should affect the budding transition when $1/m \lesssim (\kappa/\Sigma_{\alpha\beta})^{1/2}$.

This expectation is confirmed by our preliminary calculations, see Fig. \ref{Fig6}. Indeed, for $m > (\Sigma_{\alpha\beta}/\kappa)^{1/2}$, the resulting bud no longer has a spherical shape, but a tube-like structure. In fact, we find a variety of solutions to the differential equations presented in section 2. Detailed analysis on the global minimum configuration for a given liquid volume is outside the scope of this paper and will be presented elsewhere. Nonetheless, it is interesting to note that we find tube-like membrane structures even in the absence of external pulling forces, as in Ref. \cite{LiPNAS}. The tubes are initiated by the minimization of the interfacial area $A_{\alpha\beta}$ of the aqueous phases and their narrow structures are maintained by the spontaneous curvature of the membrane.

\section{Conclusions.}
In summary, partial wetting of membranes by liquid droplets leads to a new budding mechanism which is governed by the competition between bending and surface energies. Budding can be induced by increasing the volume of the liquid droplet, decreasing the membrane tension, and/or lowering the intrinsic contact angle of the droplet at the membrane. Upon performing one of these changes, the membrane becomes more strongly bent until it undergoes a discontinuous transition to a completely wrapped configuration.

Our theoretical results are accessible to experimental studies. One example is provided by micropipette experiments, in which the budding transition may be induced by varying the membrane tension. Alternatively, by deflating the vesicles, a number of relevant parameters such as droplets volume, membrane tension and intrinsic contact angle can be altered at the same time to induce budding.

In the experiments performed so far on these systems, the sizes of the two liquid compartments were comparable, whereas in our calculations, one phase is much larger than the other. It will be of great interest to extend our study and include finite size effects arising from the finite volume of the $\beta$ phase. We expect that global constraints act to suppress budding and only small buds will be possible. In this sense, the droplet-induced budding mechanism considered in this paper represents an important limiting behaviour. Other future areas of interest include a detailed analysis of the effects arising from sponatenous curvature, and the influence of the wetting properties on the neck structure of the buds.

{\it{Acknowledgements}}
We thank Yanhong Li, Yonggang Liu, and Rumiana Dimova for fruitful discussions.

\footnotesize{
\bibliography{rsc} 

\providecommand*{\mcitethebibliography}{\thebibliography}
\csname @ifundefined\endcsname{endmcitethebibliography}
{\let\endmcitethebibliography\endthebibliography}{}
\begin{mcitethebibliography}{25}
\providecommand*{\natexlab}[1]{#1}
\providecommand*{\mciteSetBstSublistMode}[1]{}
\providecommand*{\mciteSetBstMaxWidthForm}[2]{}
\providecommand*{\mciteBstWouldAddEndPuncttrue}
  {\def\EndOfBibitem{\unskip.}}
\providecommand*{\mciteBstWouldAddEndPunctfalse}
  {\let\EndOfBibitem\relax}
\providecommand*{\mciteSetBstMidEndSepPunct}[3]{}
\providecommand*{\mciteSetBstSublistLabelBeginEnd}[3]{}
\providecommand*{\EndOfBibitem}{}
\mciteSetBstSublistMode{f}
\mciteSetBstMaxWidthForm{subitem}
{(\emph{\alph{mcitesubitemcount}})}
\mciteSetBstSublistLabelBeginEnd{\mcitemaxwidthsubitemform\space}
{\relax}{\relax}

\bibitem[Li \emph{et~al.}(2008)Li, Lipowsky, and Dimova]{Li}
Y.~Li, R.~Lipowsky and R.~Dimova, \emph{{J}. {A}m. {C}hem. {S}oc.}, 2008,
  \textbf{130}, 12252\relax
\mciteBstWouldAddEndPuncttrue
\mciteSetBstMidEndSepPunct{\mcitedefaultmidpunct}
{\mcitedefaultendpunct}{\mcitedefaultseppunct}\relax
\EndOfBibitem
\bibitem[Long \emph{et~al.}(2008)Long, Cans, and Keating]{Long}
M.~S. Long, A.~S. Cans and C.~D. Keating, \emph{{J}. {A}m. {C}hem. {S}oc.},
  2008, \textbf{130}, 756\relax
\mciteBstWouldAddEndPuncttrue
\mciteSetBstMidEndSepPunct{\mcitedefaultmidpunct}
{\mcitedefaultendpunct}{\mcitedefaultseppunct}\relax
\EndOfBibitem
\bibitem[Li \emph{et~al.}(2011)Li, Lipowsky, and Dimova]{LiPNAS}
Y.~Li, R.~Lipowsky and R.~Dimova, \emph{{P}roc. {N}atl. {A}cad. {S}ci.}, 2011,
  \textbf{108}, 4731\relax
\mciteBstWouldAddEndPuncttrue
\mciteSetBstMidEndSepPunct{\mcitedefaultmidpunct}
{\mcitedefaultendpunct}{\mcitedefaultseppunct}\relax
\EndOfBibitem
\bibitem[Dominak \emph{et~al.}(2008)Dominak, Gundermann, and Keating]{Dominak}
L.~M. Dominak, E.~L. Gundermann and C.~D. Keating, \emph{{L}angmuir}, 2008,
  \textbf{130}, 756\relax
\mciteBstWouldAddEndPuncttrue
\mciteSetBstMidEndSepPunct{\mcitedefaultmidpunct}
{\mcitedefaultendpunct}{\mcitedefaultseppunct}\relax
\EndOfBibitem
\bibitem[Alberts \emph{et~al.}(1988)Alberts, Bray, Lewis, Raff, Roberts, and
  Watson]{transport}
B.~Alberts, D.~Bray, J.~Lewis, M.~Raff, K.~Roberts and J.~D. Watson,
  \emph{{M}olecular {B}iology of the {C}ell}, {G}arland {P}ublishing, {N}ew
  {Y}ork, 1988\relax
\mciteBstWouldAddEndPuncttrue
\mciteSetBstMidEndSepPunct{\mcitedefaultmidpunct}
{\mcitedefaultendpunct}{\mcitedefaultseppunct}\relax
\EndOfBibitem
\bibitem[Garoff \emph{et~al.}(1998)Garoff, Hewson, and Opstelten]{viral}
H.~Garoff, R.~Hewson and D.~J.~E. Opstelten, \emph{{M}icrobiol. {M}ol. {B}iol.
  {R}ev.}, 1998, \textbf{62}, 1171\relax
\mciteBstWouldAddEndPuncttrue
\mciteSetBstMidEndSepPunct{\mcitedefaultmidpunct}
{\mcitedefaultendpunct}{\mcitedefaultseppunct}\relax
\EndOfBibitem
\bibitem[Discher and Eisenberg(2002)]{Discher}
D.~E. Discher and A.~Eisenberg, \emph{{S}cience}, 2002, \textbf{297},
  967--973\relax
\mciteBstWouldAddEndPuncttrue
\mciteSetBstMidEndSepPunct{\mcitedefaultmidpunct}
{\mcitedefaultendpunct}{\mcitedefaultseppunct}\relax
\EndOfBibitem
\bibitem[Karlsson \emph{et~al.}(2001)Karlsson, Karlsson, M.~Karlsson,
  Str\"{o}mberg, Rytts\'{e}n, and Orwar]{Orwar}
A.~Karlsson, R.~Karlsson, A.-S.~C. M.~Karlsson, A.~Str\"{o}mberg,
  F.~Rytts\'{e}n and O.~Orwar, \emph{{N}ature}, 2001, \textbf{409},
  150--152\relax
\mciteBstWouldAddEndPuncttrue
\mciteSetBstMidEndSepPunct{\mcitedefaultmidpunct}
{\mcitedefaultendpunct}{\mcitedefaultseppunct}\relax
\EndOfBibitem
\bibitem[Noireaux and Libchaber(2004)]{Libchaber}
V.~Noireaux and A.~Libchaber, \emph{{P}roc. {N}atl. {A}cad. {S}ci.}, 2004,
  \textbf{101}, 17669--17674\relax
\mciteBstWouldAddEndPuncttrue
\mciteSetBstMidEndSepPunct{\mcitedefaultmidpunct}
{\mcitedefaultendpunct}{\mcitedefaultseppunct}\relax
\EndOfBibitem
\bibitem[Lipowsky(1992)]{Lipowsky}
R.~Lipowsky, \emph{{J}. {P}hys. {II}}, 1992, \textbf{2}, 1825\relax
\mciteBstWouldAddEndPuncttrue
\mciteSetBstMidEndSepPunct{\mcitedefaultmidpunct}
{\mcitedefaultendpunct}{\mcitedefaultseppunct}\relax
\EndOfBibitem
\bibitem[J\"ulicher and Lipowsky(1996)]{Julicher}
F.~J\"ulicher and R.~Lipowsky, \emph{{P}hys. {R}ev. {E}}, 1996, \textbf{54},
  2670\relax
\mciteBstWouldAddEndPuncttrue
\mciteSetBstMidEndSepPunct{\mcitedefaultmidpunct}
{\mcitedefaultendpunct}{\mcitedefaultseppunct}\relax
\EndOfBibitem
\bibitem[Baumgart \emph{et~al.}(2003)Baumgart, Hess, and Webb]{Baumgart}
T.~Baumgart, S.~T. Hess and W.~W. Webb, \emph{{N}ature}, 2003, \textbf{425},
  821\relax
\mciteBstWouldAddEndPuncttrue
\mciteSetBstMidEndSepPunct{\mcitedefaultmidpunct}
{\mcitedefaultendpunct}{\mcitedefaultseppunct}\relax
\EndOfBibitem
\bibitem[Baumgart \emph{et~al.}(2005)Baumgart, Das, Webb, and Jenkins]{Baumg}
T.~Baumgart, S.~Das, W.~W. Webb and J.~T. Jenkins, \emph{{B}iophys. {J}.},
  2005, \textbf{89}, 1067--1080\relax
\mciteBstWouldAddEndPuncttrue
\mciteSetBstMidEndSepPunct{\mcitedefaultmidpunct}
{\mcitedefaultendpunct}{\mcitedefaultseppunct}\relax
\EndOfBibitem
\bibitem[Bacia \emph{et~al.}(2005)Bacia, Schwille, and Kurzchalia]{Bacia}
K.~Bacia, P.~Schwille and T.~Kurzchalia, \emph{{P}roc. {N}atl. {A}cad. {S}ci.},
  2005, \textbf{102}, 3272--3277\relax
\mciteBstWouldAddEndPuncttrue
\mciteSetBstMidEndSepPunct{\mcitedefaultmidpunct}
{\mcitedefaultendpunct}{\mcitedefaultseppunct}\relax
\EndOfBibitem
\bibitem[Dimova \emph{et~al.}(2007)Dimova, Riske, Aranda, Bezlyepkina, Knorr,
  and Lipowsky]{Dimova}
R.~Dimova, K.~A. Riske, S.~Aranda, N.~Bezlyepkina, R.~L. Knorr and R.~Lipowsky,
  \emph{{S}oft {M}atter}, 2007, \textbf{3}, 817--827\relax
\mciteBstWouldAddEndPuncttrue
\mciteSetBstMidEndSepPunct{\mcitedefaultmidpunct}
{\mcitedefaultendpunct}{\mcitedefaultseppunct}\relax
\EndOfBibitem
\bibitem[Semrau \emph{et~al.}(2008)Semrau, Idema, Holtzer, Schmidt, and
  Storm]{Semrau}
S.~Semrau, T.~Idema, L.~Holtzer, T.~Schmidt and C.~Storm, \emph{{P}hys. {R}ev.
  {L}ett.}, 2008, \textbf{100}, 088101\relax
\mciteBstWouldAddEndPuncttrue
\mciteSetBstMidEndSepPunct{\mcitedefaultmidpunct}
{\mcitedefaultendpunct}{\mcitedefaultseppunct}\relax
\EndOfBibitem
\bibitem[McMahon and Gallop(2005)]{McMahon}
H.~T. McMahon and J.~L. Gallop, \emph{{N}ature}, 2005, \textbf{438}, 590\relax
\mciteBstWouldAddEndPuncttrue
\mciteSetBstMidEndSepPunct{\mcitedefaultmidpunct}
{\mcitedefaultendpunct}{\mcitedefaultseppunct}\relax
\EndOfBibitem
\bibitem[Hurley \emph{et~al.}(2010)Hurley, Boura, Carlson, and
  R\'{o}\.{z}ycki]{Hurley}
J.~H. Hurley, E.~Boura, L.-A. Carlson and B.~R\'{o}\.{z}ycki, \emph{{C}ell},
  2010, \textbf{143}, 875\relax
\mciteBstWouldAddEndPuncttrue
\mciteSetBstMidEndSepPunct{\mcitedefaultmidpunct}
{\mcitedefaultendpunct}{\mcitedefaultseppunct}\relax
\EndOfBibitem
\bibitem[Lipowsky and D\"{o}bereiner(1998)]{Lipowsky2}
R.~Lipowsky and H.-G. D\"{o}bereiner, \emph{{E}urophys. {L}ett.}, 1998,
  \textbf{43}, 219\relax
\mciteBstWouldAddEndPuncttrue
\mciteSetBstMidEndSepPunct{\mcitedefaultmidpunct}
{\mcitedefaultendpunct}{\mcitedefaultseppunct}\relax
\EndOfBibitem
\bibitem[Deserno(2004)]{Deserno}
M.~Deserno, \emph{{P}hys. {R}ev. {E}}, 2004, \textbf{69}, 031903\relax
\mciteBstWouldAddEndPuncttrue
\mciteSetBstMidEndSepPunct{\mcitedefaultmidpunct}
{\mcitedefaultendpunct}{\mcitedefaultseppunct}\relax
\EndOfBibitem
\bibitem[Nowak and Chou(2008)]{Nowak}
S.~A. Nowak and T.~Chou, \emph{{P}hys. {R}ev. {E}}, 2008, \textbf{78},
  021908\relax
\mciteBstWouldAddEndPuncttrue
\mciteSetBstMidEndSepPunct{\mcitedefaultmidpunct}
{\mcitedefaultendpunct}{\mcitedefaultseppunct}\relax
\EndOfBibitem
\bibitem[Kusumaatmaja \emph{et~al.}(2009)Kusumaatmaja, Li, Dimova, and
  Lipowsky]{Kusumaatmaja}
H.~Kusumaatmaja, Y.~Li, R.~Dimova and R.~Lipowsky, \emph{{P}hys. {R}ev.
  {L}ett.}, 2009, \textbf{103}, 238103\relax
\mciteBstWouldAddEndPuncttrue
\mciteSetBstMidEndSepPunct{\mcitedefaultmidpunct}
{\mcitedefaultendpunct}{\mcitedefaultseppunct}\relax
\EndOfBibitem
\bibitem[Goetz and Lipowsky(1998)]{Goetz}
R.~Goetz and R.~Lipowsky, \emph{{J}. {C}hem. {P}hys.}, 1998, \textbf{108},
  7397\relax
\mciteBstWouldAddEndPuncttrue
\mciteSetBstMidEndSepPunct{\mcitedefaultmidpunct}
{\mcitedefaultendpunct}{\mcitedefaultseppunct}\relax
\EndOfBibitem
\bibitem[Press \emph{et~al.}(1992)Press, Flannery, Teukolsky, and
  Vetterling]{RungeKutta}
W.~H. Press, B.~P. Flannery, S.~A. Teukolsky and W.~T. Vetterling,
  \emph{{N}umerical {R}ecipes in {C}}, {C}ambridge {U}niversity {P}ress,
  {C}ambridge, 1992\relax
\mciteBstWouldAddEndPuncttrue
\mciteSetBstMidEndSepPunct{\mcitedefaultmidpunct}
{\mcitedefaultendpunct}{\mcitedefaultseppunct}\relax
\EndOfBibitem
\bibitem[Rowlinson and Widom(2003)]{Widom}
J.~S. Rowlinson and B.~Widom, \emph{{M}olecular {T}heory of {C}apillarity},
  {D}over, 2003\relax
\mciteBstWouldAddEndPuncttrue
\mciteSetBstMidEndSepPunct{\mcitedefaultmidpunct}
{\mcitedefaultendpunct}{\mcitedefaultseppunct}\relax
\EndOfBibitem
\end{mcitethebibliography}
\bibliographystyle{rsc} 
}

\end{document}